\documentclass[RNAAS]{aastex63}

\usepackage{xspace}
\newcommand{\name}{SN~2023ixf\xspace}

\begin{document}

\title{No UV-bright Eruptions from SN~2023ixf in GALEX Imaging 15--20 Years Before Explosion}

\correspondingauthor{Nicholas Flinner}
\email{nicholasflinner@gmail.com}

\author[0009-0009-9597-8565]{Nicholas Flinner}
\affiliation{Department of Astronomy, The Ohio State University, 140 West 18th Avenue, Columbus, OH, USA}

\author[0000-0002-2471-8442]{Michael A. Tucker}
\altaffiliation{CCAPP Fellow}
\affiliation{Department of Astronomy, The Ohio State University, 140 West 18th Avenue, Columbus, OH, USA}
\affiliation{Department of Physics, The Ohio State University, 191 West Woodruff Ave, Columbus, OH, USA}
\affiliation{Center for Cosmology and Astroparticle Physics, The Ohio State University, 191 West Woodruff Ave, Columbus, OH, USA}

\author{John F. Beacom}
\affiliation{Department of Astronomy, The Ohio State University, 140 West 18th Avenue, Columbus, OH, USA}
\affiliation{Department of Physics, The Ohio State University, 191 West Woodruff Ave, Columbus, OH, USA}
\affiliation{Center for Cosmology and Astroparticle Physics, The Ohio State University, 191 West Woodruff Ave, Columbus, OH, USA}

\author{Benjamin J. Shappee}
\affiliation{Institute for Astronomy, University of Hawai‘i at Mānoa, 2680 Woodlawn Dr., Honolulu, HI, USA}

\begin{abstract}

We analyze pre-explosion ultraviolet (UV) imaging of the nearby Type II supernova \name in search of precursor variability. No outbursts are seen in observations obtained 15--20~yr prior to explosion to a limit of $L_{\rm NUV} \approx 1000~L_\odot$ and $L_{\rm FUV} \approx 2000~L_\odot$. The time period of these non-detections roughly corresponds to changes in the circumstellar density inferred from early spectra and photometry.
 
\end{abstract}

\section{Introduction} \label{sec:intro}

Core-collapse supernovae (CCSNe) are the birthplace of black holes and neutron stars (e.g., \citealp{wilson71, foglizzo15}). They originate from massive ($\gtrsim 8~M_{\odot}$) stars, but the stellar evolution in the decades prior to explosions remains unclear. There is some observational evidence for enhanced mass-loss prior to explosion traced by narrow emission lines (e.g., \citealp{sollerman20}), peculiar photometric evolution (e.g., \citealp{purisiainen23}), or both (e.g., \citealp{nyholm17}). However, there is also evidence that pre-explosion variability does not occur \citep{johnson18}. Even if it occurs, the physical mechanisms are not well established \citep{smith14}. Some models invoke gravity waves to deposit energy into the stellar envelope (e.g., \citealp{quataert12}); others rely on impulsive energy injection from the nuclear ignition of different elements (e.g., \citealp{woosley15}). Each theory predicts differences in the amount, duration, and onset of the increased mass-loss phase.

Photometry of the progenitor star in the years prior to explosion constrains the production of circumstellar material (CSM; e.g., \citealp{khatami23}). The luminous blue variable progenitor of SN~2009ip experienced multiple eruptive episodes in the decade prior to explosion (e.g., \citealp{smith10}). Conversely, the red supergiant (RSG) progenitors of SNe~2020tlf \citep{jacobson22} and 2021qqp \citep{hiramastsu23} showed steady increases in brightness $\lesssim 2~\rm yr$ before explosion. This hints at different underlying processes governing CSM production. It remains unclear if enhanced mass-loss is common to all SNe~II or if the few observed events constitute a special class.

\citet{itagaki23} discovered \name on 2023-05-19 (MJD 60083) in the nearby face-on spiral galaxy M101 \citep[$z = 0.000804$;][]{vaucouleurs95} located $d=6.4~\rm Mpc$ \citep{shappee11} away. It was classified as a Type II by \citet{perley23}. Its proximity allows for detailed study of its pre- and post-explosion properties.

\section{GALEX Data and Analysis} \label{sec:methods}

The Galaxy Evolution Explorer \citep[GALEX;][]{martin05, morrissey07} space telescope operated from 2003--2013 in two ultraviolet (UV) filters: the near-UV (NUV, ${\lambda_{\rm eff} = 2315.7}$~\AA) and the far-UV (FUV, ${\lambda_{\rm eff} = 1538.6}$~\AA). GALEX has a total of $\approx 15,000~\rm{s}$ of simultaneous exposure time in both filters at the site of \name. We queried the GALEX photometry using gPhoton \citep{million16} with a $6\arcsec$ aperture and 5 identical apertures placed on nearby blank regions to estimate and subtract the image background (Fig. 1). 

We detect statistically significant NUV flux at the location of \name, but inspection of the images reveals part of an H~II region within the aperture. The FUV observations produce only non-detections (signal/noise $<3\sigma$). The average fluxes within the apertures are ${f_{\rm NUV} = (6.6 \pm 0.4) \times 10^{-16}~\rm \; erg\; cm^{-2}\; s^{-1}}$~\AA$^{-1}$ and ${f_{\rm FUV} = (1.1 \pm 0.2) \times 10^{-15}~\rm\; erg\; cm^{-2}\; s^{-1}}$~\AA$^{-1}$. The flux uncertainties are validated by performing identical analyses on nearby blank regions. The GALEX data thus shows no evidence for precursor UV variability in \name from 2003--2008.

\begin{figure*}
    \centering
    \includegraphics[width=0.95\linewidth]{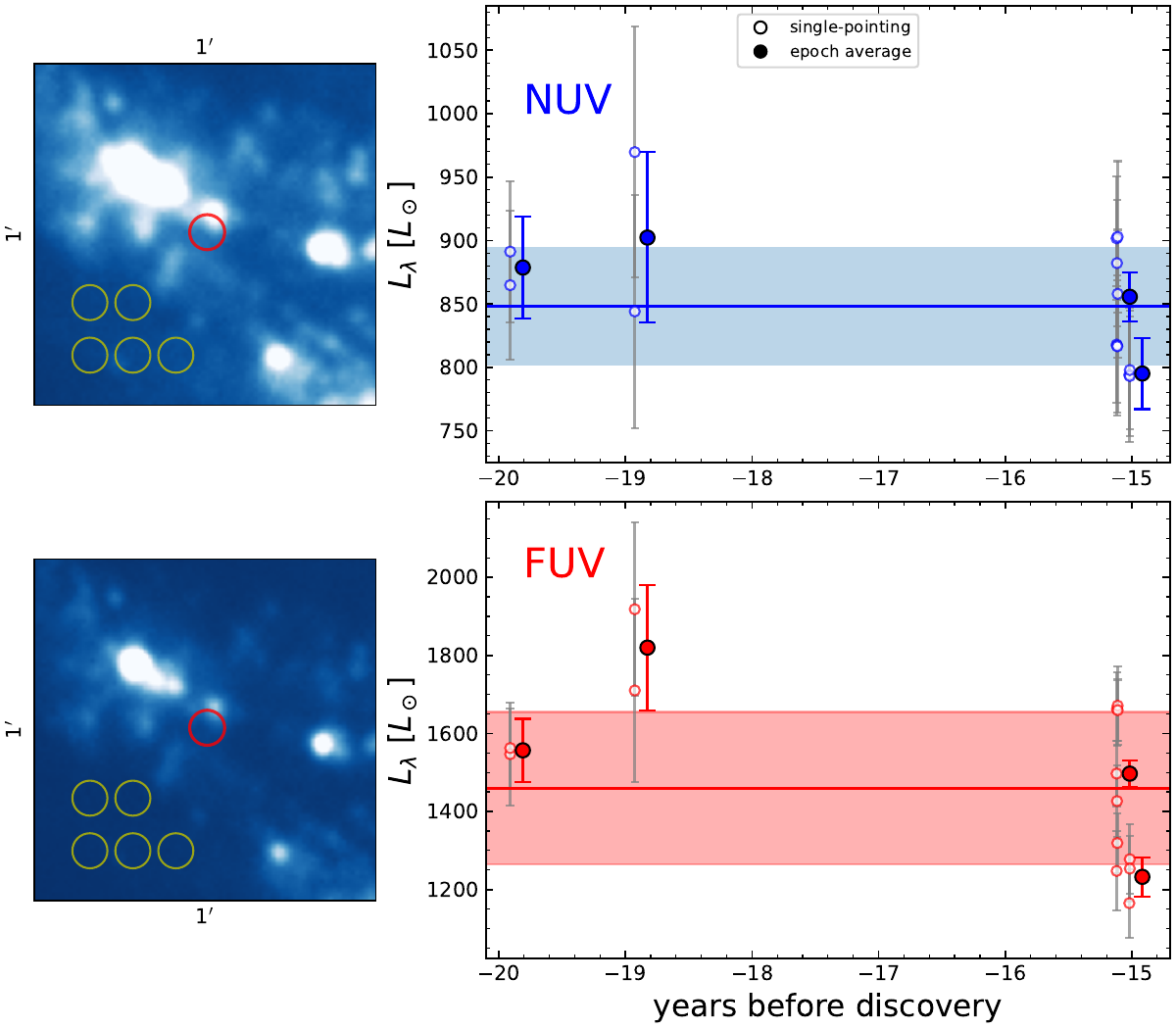}
    \caption{\textit{Left Panels}: $1 \arcmin \times 1 \arcmin$ cutouts centered on the location of \name. Insets are shown for both filters and are averaged over all $\approx 15,000~\rm s$ of exposure time. The red circle represent the aperture and the yellow circles represent the background.
        \textit{Right Panels}: NUV (top) and FUV (bottom) light curves of single-pointing (open symbols) and the weighted averages of observations separated by $> 0.1~\rm yr$ and shifted 0.1~yr for visual clarity (filled symbols). The solid lines are the weighted average of all single-pointing luminosities and the shaded bands represent $1\sigma$ scatter. 
        }
\label{fig:1}
\end{figure*}

\section{Discussion} \label{sec:discuss}

After correcting for foreground Milky Way extinction \citep{schlafy11} and computing the scatter from individual GALEX measurements,\footnote{We neglect uncertainties in the distance to M101 in our variability analysis.} 
we find no UV-bright outbursts or eruptions to a limit of $L_{\rm NUV} \approx 1000~L_\odot$ and $L_{\rm FUV} \approx 2000~L_\odot$ during the GALEX observations. Multiple studies have analyzed the early light curves and spectra of \name (e.g., \citealp{jacobsongalan23, grefenstette23, smith23}) finding a relatively normal SN~II. Intriguingly, a consistent claim is a decrease in the CSM density at $r \approx 10^{15}~\rm cm$. The ejecta velocity of $v_{\rm ej} \approx 5,400~\rm km/s$ derived by \citet{grefenstette23} corresponds to a shift in CSM production from the progenitor about two decades prior to explosion \citep{singhteja23}.

The progenitor for \name is inferred to be a near/mid-IR variable RSG shrouded in $A_V \approx 4.6 \pm 0.2~\rm mag$ of dust \citep{pledger23, kilpatrick23, jencson23} that likely produces negligible UV emission. The dusty shell will similarly suppress any UV photons produced by precursor activity but it's difficult to hide eruptions across all wavelengths (see, e.g., \citealp{neustadt23}). Other RSGs with precursor emission show steady rises prior to explosion, suggestive of enhanced winds rather than eruptive phenomena.

This analysis extends the non-detections of precursor emission in the optical \citep{neustadt23, dong23} to the UV and even earlier epochs. The GALEX observations occurred around the time that the progenitor of \name entered the enhanced mass-loss phase, yet we see no evidence for variability. While the dust surrounding the star naturally shifts energy to longer wavelengths, future modeling efforts of the pre-explosion evolution must ensure that the CSM production mechanism remains faint and/or rapid at UV wavelengths to avoid tensions with these observations.

\section*{Acknowledgements}

We thank C. S. Kochanek for useful discussions. NLF was supported by the Robert P. Caren Family Endowment Fund. JFB was supported by NSF Grant No.\ PHY-2012955. 

\bibliography{2023ixf}{}
\bibliographystyle{aasjournal}
\end{document}